\documentclass[journal,twoside,web]{ieeecolor}
\usepackage{generic}
\usepackage{cite}
\usepackage{amsmath,amssymb,amsfonts}
\usepackage{graphicx}
\usepackage{algorithm,algorithmic}
\usepackage{hyperref}
\usepackage{textcomp}
\usepackage{dblfloatfix}
\usepackage{mathtools}
\usepackage{algorithmic}
\usepackage{balance}
\usepackage{mathrsfs}
\usepackage[switch]{lineno}
\usepackage{booktabs}
\usepackage{tikz}
\definecolor{myblue}{rgb}{0, 0.447, 0.741}
\usepackage{siunitx}
\usetikzlibrary{arrows.meta,positioning,calc,patterns,decorations.pathmorphing}
\usepackage{makecell}

\newtheorem{remm}{Remark}

\newtheorem{examp}{Example}

\newtheorem{thm}{Theorem}
\newenvironment{theorem}{\begin{thm}\rm }{\hfill \hspace*{1pt} \hfill $\lrcorner$ \end{thm}}

\newtheorem{assumption}{Assumption}
\newenvironment{assume}{\begin{assumption}\rm }{\hfill \hspace*{1pt} \hfill $\lrcorner$ \end{assumption}}

\newtheorem{defi}{Definition}
\newenvironment{definition}[1]{\begin{defi}[#1]\rm }{\hfill \hspace*{1pt} \hfill $\lrcorner$ \end{defi}}

\def\BibTeX{{\rm B\kern-.05em{\sc i\kern-.025em b}\kern-.08em
    T\kern-.1667em\lower.7ex\hbox{E}\kern-.125emX}}
\begin{document}

\title{Passive Lifted FIR Filters  for \\ Nonlinear System Identification}
\author{Zixing Wang$^{a}$, \IEEEmembership{Graduate Student Member, IEEE}, Fulvio Forni$^{a}$, \IEEEmembership{Senior Member, IEEE}
\thanks{The work of Zixing Wang is supported by CSC Cambridge Scholarship and Robinson College, Cambridge. }
\thanks{$^{a}$ Department of Engineering, University of Cambridge, CB2 1PZ Cambridge, U.K.
         \{{\tt\small zxw20@cam.ac.uk}, {\tt\small f.forni@eng.cam.ac.uk}\}.}
}

\maketitle

\begin{abstract}
Passivity is a fundamental property of physical systems. In data-driven modeling, ensuring that a learned model preserves this structural property is critical to avoiding instability in close loop. Although linear passive system identification is well-established, nonlinear extensions remain challenging. We propose nonlinear operators defined through passivity-preserving lifting of linear passive FIR filters. Passivity is enforced efficiently through frequency-domain constraints, and the nonlinear lifting includes output feedback for expressivity. Numerical \textcolor{black}{and real-world} experiments demonstrate the framework capabilities, including the computational advantage of frequency-domain constraints against LMI-based alternatives.
\end{abstract}


\section{Introduction} \label{sec:introduction}
Passivity is a common property of most physical systems \cite{Willems1972, Ortega1998eulerbook, Schaft1999}. It formalizes the idea that if a physical system has no internal sources, its internal stored energy must be bounded by the energy supplied at its ports \cite{Ortega1998eulerbook, Schaft1999}. For feedback systems, the passivity theorem guarantees that the feedback interconnection of passive systems remains passive \cite{Willems1972, HillANDMoylan1977, Rodolphe1996}. This makes passivity a crucial compositional property in the analysis and synthesis of interconnected systems. 

In system identification and machine learning, it is crucial that any learned model of a passive system is guaranteed to be passive by construction. We argue that the structural, physical nature of passivity calls for learned models to be passive regardless of the approximation error between the system and the model. This is because passivity enables stronger robustness properties, such as the infinite gain margin observed in linear systems, thereby providing more rigorous guarantees for controller synthesis. 

Consider the identification of an electromechanical system for the design or training of a feedback controller in simulation. If the learned model fails to preserve the passivity of the physical system, the simulated closed-loop response may diverge significantly from the actual system behavior \cite{Ye2011}. Specifically, a non-passive learned model interconnected in negative feedback with a passive controller may exhibit artificial closed-loop instability, even when the underlying physics dictates that such instability is physically impossible. This discrepancy fundamentally undermines
the very purpose of learning a model for control synthesis purposes.

Although linear passive system identification is well developed \cite{Khosravi_Smith2023}, progress in the nonlinear setting remains relatively sparse and may be broadly divided into two categories: state-space approaches and input-output approaches. In the state-space setting, works such as \cite{Cheng2016,Smith2024,Manchester2024} enforce passivity by imposing Lagrangian or Hamiltonian structure, with models learned using kernel methods or neural networks. Other state-space approaches seek to adapt passivity constraints originally developed for LTI systems, for example by expressing the model in a Koopman operator framework \cite{HIS2020} or an LPV \textcolor{black}{framework~\cite{Toth2026}}. 
In the input-output setting, the state of the art includes 
the incrementally passive nonlinear kernel framework
of \cite{Henk2023} and \cite{Tony2024}. The relaxation to (non-incremental) passivity properties remains challenging \cite[Sec. I]{Henk2024}. A solution is proposed in \cite{Henk2024}, taking advantage of nonlinear kernel modeling  with passivity enforced through linear
matrix inequalities (LMIs). However, the lack of causality and the computational costs of LMIs for large data are significant bottlenecks in applications.

In this paper, we look into the problem of passive nonlinear system identification in the input-output setting. We propose a new Single-Input Single-Output (SISO) model structure based on a particular family of Nonlinear Finite Impulse Response (NFIR) operators.  Our approach leverages a passivity-preserving lifting of linear passive FIR filters into a larger feature space. This involves the pre- and post-composition with a ‘lifting’ operator and its adjoint, both carefully designed to ensure the causality, passivity and $\ell_{2}$ stability of the resulting nonlinear operator. We learn both FIR filters and nonlinear lifting operators from input/output data, via alternating optimization.

Compared with existing nonlinear approaches, the distinctive feature of our NFIR operator is the use of  a lifting operator that can accommodate output feedback. This extends the feedforward framework of \cite{Henk2024} to a recursive setting. We hypothesize that this recursive lifting enhances the model's expressivity and approximation capabilities, a claim we validate through numerical experiments. 
The lifting operator also allows classical passivity constraints for LTI systems to be directly applied to nonlinear system identification. This includes both linear matrix inequalities and frequency-domain conditions. For the frequency domain, 
we prove that a conservative positive realness condition combined with finite sampling is sufficient to guarantee the passivity of the NFIR operator. This frequency-domain approach produces a set of linear constraints that are significantly more tractable than the classical LMI-based constraints used in the passive Koopman operator framework \cite{HIS2020}. Our experiments demonstrate that these frequency-domain constraints can be solved orders of magnitude faster than their LMI-based counterparts.

This paper is structured as follows. Section~\ref{sec:lifting-based passive operators} introduces the passivity-preserving lifting mechanism at the core of NFIR operators. Through this mechanism, any passive linear operator can be lifted to a passive nonlinear operator. NFIR operators are introduced in Section~\ref{sec:imple_operator} by restricting linear filters to FIR filters and the lifting mechanism to a finite representation. 
Learning of NFIR operators is discussed in Section~\ref{sec:freq_vs_time}, with further details on hyper-parameters and implementation of lifting operators in Section~\ref{sec:hyper_lift}. 
The method is illustrated in Section~\ref{sec:examples} through examples. A brief comparison with the literature and concluding remarks complete the paper.

\section{Passive lifted operators} \label{sec:lifting-based passive operators}

\subsection{Mathematical preliminaries} 
\label{sec:preli}
We consider the Hilbert space $\ell_{2}^{m}$ given by the set of functions $u:\mathbb{Z} \to \mathbb{R}^{m}$ such that \textcolor{black}{$\sum_{t\in \mathbb{Z}} \left\Vert u(t) \right\Vert_{2}^{2} < \infty $}. The time axis $\mathbb{Z}$ is the set of integers. $\langle\cdot,\cdot \rangle_{\ell_{2}^{m}}$ is the standard inner product on $\ell_{2}^{m}$, given by $\langle u, y \rangle_{\ell_{2}^{m}}= 
 \sum_{t\in \mathbb{Z}} u(t)^{T}y(t)$ for all $u,y \in \ell_{2}^{m}$. The induced norm reads $\lVert \cdot \rVert_{\ell_{2}^{m}} = \sqrt{\langle\cdot,\cdot \rangle_{\ell_{2}^{m}}}$. For brevity, we denote $\ell_{2}^{1}$ as $\ell_{2}$, and write $\langle\cdot,\cdot \rangle_{\ell_{2}^{m}}$ as $\langle\cdot,\cdot \rangle$ when there is no ambiguity.  

We denote the Banach space of bounded linear operators from  $\ell_{2}^{m_{1}}$ to $\ell_{2}^{m_{2}}$ as
\begin{align}
    &\mathbb{B}(\ell_{2}^{m_{1}},\ell_{2}^{m_{2}}):= \{ \mathcal{G}: \ell_{2}^{m_{1}} \to \ell_{2}^{m_{2} } \,|\, \mathcal{G} \text{ is linear}  \nonumber \\
    & \text{ and } \exists c\geq0  \text{ s.t. } \Vert \mathcal{G}u \Vert_{\ell_{2}^{m_{2}}} \leq c \Vert u \Vert_{\ell_{2}^{m_{1}}} \quad  \forall u \in \ell_{2}^{m_{1}}\}.
\end{align}
For all $\mathcal{G} \in \mathbb{B}(\ell_{2}^{m_{1}},\ell_{2}^{m_{2}})$, the operator norm is defined as $\Vert \mathcal{G}\Vert  := \sup_{u \neq 0}\Vert \mathcal{G}u \Vert_{\ell_{2}^{m_{2}}}/\Vert u \Vert_{\ell_{2}^{m_{1}}}$. We define $\mathcal{G}^{\dagger} \in \mathbb{B}(\ell_{2}^{m_{2}},\ell_{2}^{m_{1}})$ as its adjoint so that for all $u \in \ell_{2}^{m_{1}}$ and all $v \in \ell_{2}^{m_{2}}$, we have $\langle\mathcal{G}u, v \rangle_{\ell_{2}^{m_{2}}} = \langle u, \mathcal{G}^{\dagger}v \rangle_{\ell_{2}^{m_{1}}}$.

Given any positive integer $R$, 
we define the \emph{projection} operator $\mathcal{P}_{\tau}^{R}:\ell_{2}^{m} \to \ell_{2}^{m}$ as
\begin{equation} \label{eq:P_tau_R}
    (\mathcal{P}_{\tau}^{R} u)(t) 
    = 
   \begin{cases}
    \textcolor{black}{u(t)  \quad \tau-R+1 \leq t \leq\tau } \\
    0 \quad\quad \text{otherwise}.
  \end{cases}
\end{equation}
for all $\tau \in \mathbb{Z}$ and all $u \in \ell_{2}^{m}$. 
We will adopt the notation $\mathcal{P}_{\tau}:\ell_{2}^{m} \to \ell_{2}^{m}$ 
for the case $R=\infty$, that is, 
\begin{equation} \label{eq:P_tau}
    (\mathcal{P}_{\tau} u)(t) 
    = 
   \begin{cases}
    u(t) \quad t \leq \tau \\
    0 \quad \quad  t > \tau.
  \end{cases}
\end{equation}

In a similar way, given any positive integer $R$, 
we define the \emph{truncation} operator $\bar{\mathcal{P}}_{\tau}^{R}:\ell_{2}^{m} \to \ell_{2[0,R-1]}^{m}$ as
\begin{equation} \label{eq:P_tau_R_bar}
    (\bar{\mathcal{P}}_{\tau}^{R} u)(t) 
    = 
    \textcolor{black}{u(\tau - R + 1 + t) \quad t \in [0,R-1] \cap \mathbb{Z}} 
\end{equation}
for all $\tau \in \mathbb{Z}$ and all $u \in \ell_{2}^{m}$.

An operator $\mathcal{G}:\ell_{2}^{m_{1}} \to \ell_{2}^{m_{2}} $ is \emph{causal} (respectively, \emph{strictly causal} )\cite{Henk2024,desoer_Vidyaasagar} if for all $u\in \ell_{2}^{m_{1}}$ and for all $\tau \in \mathbb{Z},  \mathcal{P}_{\tau}\mathcal{G}u = \mathcal{P}_{\tau}\mathcal{G}\mathcal{P}_{\tau}u$ (respectively, $\mathcal{P}_{\tau}\mathcal{G}u = \mathcal{P}_{\tau}\mathcal{G}\mathcal{P}_{\tau -1}u$).
An operator $\mathcal{G}:\ell_{2}^{m} \to \ell_{2}^{m} $ is \emph{passive}\textcolor{black}{ \cite{Henk2024,desoer_Vidyaasagar}} if for all $u\in \ell_{2}^{m}$ and all $\tau \in \mathbb{Z} $, we have $\langle\mathcal{P}_{\tau}u, \mathcal{P}_{\tau}\mathcal{G}u\rangle \geq 0$.
An operator $\mathcal{G}:\ell_{2}^{m_{1}} \to \ell_{2}^{m_{2}} $ has \emph{finite $\ell_{2}$ gain} \textcolor{black}{\cite{desoer_Vidyaasagar}} less than or equal to $\gamma \geq 0$ if, for all $u\in \ell_{2}^{m_{1}}$, $\langle \mathcal{G}u,\mathcal{G}u \rangle_{\ell_{2}^{m_{2}}} \leq \gamma^{2} \langle u,u\rangle_{\ell_{2}^{m_{1}}}$.
Finally, we use \textcolor{black}{$\ell_{2[0,L-1]}^{m}$} to denote the Hilbert space of functions $u:\{0,1,\dots,L-1\}\to \mathbb{R}^{m}$ such that $\sum_{t=0}^{L-1} \Vert u(t) \Vert_{2}^{2} < \infty $. The inner product and induced norm can be defined similarly. We denote $\ell_{2[0,L-1]}^{1}$ as $\ell_{2[0,L-1]}$. The projection $\mathcal{P}_{\tau}^{R}$ can also be defined on $\ell_{2[0,L-1]}^{m}$ with little modification to \eqref{eq:P_tau_R}. 

\subsection{Lifted operator and passivity} \label{sec:lifting_operator}
\begin{definition} {Lifted nonlinear operator} \label{defi:lift_operator}

A lifted operator $\mathcal{F}$ with input $u \in \ell_{2}$,
output $y \in \ell_{2}$, and additional input $q \in \ell_{2}^{m_{q}}$ is given by
    \begin{equation} \label{eq:lift_operator}
        y = \mathcal{F}(u,q) := \mathcal{N}(u,q)^{\dagger}\mathcal {G}\mathcal{N}(u,q)u
    \end{equation}
    where
\begin{itemize}
\item 
$\mathcal{G} \in \mathbb{B}(\ell_{2}^{m_{\mathcal{N}}},\ell_{2}^{m_{\mathcal{N}}})$ is any causal, Multi-Input Multi-Output (MIMO), $\ell_{2}$ stable linear time invariant (LTI) filter;
\item 
$\mathcal{N} : \ell_{2} \times \ell_{2}^{m_{q}} \to \mathbb{B}(\ell_{2},\ell_{2}^{m_{\mathcal{N}}})$ is a causal operator. 
Furthermore, $\mathcal{N}(u,q)$ and its adjoint $\mathcal{N}(u,q)^{\dagger}$ are causal operators for all $(u,q) \in \ell_{2} \times \ell_{2}^{m_{q}}$.  
\end{itemize} \vspace{-5mm}
\end{definition}
\vspace{2mm}

The operator in Definition~\ref{defi:lift_operator} has the interpretation in Fig.~\ref{fig:lift_graph}. \textcolor{black}{The operator} $\mathcal{N}$ can be interpreted as a lifting function that lifts the input space $\ell_{2}$ to a high-dimensional space  $\ell_{2}^{m_{\mathcal{N}}}$. The lifting function itself depends on the operator input $u$
and on the additional input $q$. The linear operator $\mathcal{G}$ maps the lifted signal $w = \mathcal{N}(u,q)u$ to $\mathcal{G}w$ in the lifted space. Finally, the result is mapped back to the low-dimensional output space via the adjoint of the lifting function.
We call $q$ the \emph{additional} input, as its role is to 
add flexibility to the lifting function. It shapes the function and can be removed if necessary. For instance, $q \in \ell_{2}^{m_{q}}$ could be an  exogenous reference used to customize the lifting to a particular working point. Likewise, it could be used to introduce recursion through the strictly causal relationship $q(k) = y(k-1)$ for all $k \in \mathbb{Z}$.

\begin{figure}[htbp]
    \centering
    \begin{tikzpicture}[
        scale=0.7,
        transform shape,
        >=Stealth,
        line width=0.9pt,
        every node/.style={font=\small},
        op/.style={-{Stealth[length=2.2mm,width=1.8mm]}},
        dot/.style={circle, fill=black, inner sep=1.4pt}
    ]

        \draw (0,0) ellipse (3.1cm and 0.62cm);
        \draw (0,4.0) ellipse (5.1cm and 1.2cm);

        \node at (5.55,4.0) {$\ell_2^{m_{\mathcal{N}}}$};
        \node at (3.45,0.0) {$\ell_2$};

        \coordinate (u)  at (2.55,-0.10);
        \coordinate (w)  at (2.00,4.02);
        \coordinate (gw) at (-1.10,4.10);
        \coordinate (y)  at (-2.35,-0.12);

        \node[dot, label={[below right=-1mm and -1mm]$u$}] at (u) {};
        \node[dot, label={[above right=-1mm and -1mm]{$w=\mathcal{N}(u,q)u$}}] at (w) {};
        \node[dot, label={[above left=-1mm and -1mm]{$\mathcal{G}w$}}] at (gw) {};
        \node[dot, label={[below left=-1mm and -1mm]$y$}] at (y) {};

        \draw[op]
            (u) .. controls +(0,1.5) and +(0,-1.4) ..
            node[pos=0.52, right=8pt] {$\mathcal{N}(u,q)$}
            (w);

        \draw[op]
            (w) .. controls +(-0.75,0.70) and +(0.75,0.80) ..
            node[pos=0.48, above] {$\mathcal{G}$}
            (gw);

        \draw[op]
            (gw) .. controls +(0,-1.45) and +(0,1.50) ..
            node[pos=0.52, left=10pt] {$\mathcal{N}(u,q)^{\dagger}$}
            (y);

    \end{tikzpicture}
    \caption{A graphical interpretation of the lifted operator
    $y=\mathcal{N}(u,q)^{\dagger}\mathcal{G}\mathcal{N}(u,q)u$.}
    \label{fig:lift_graph}
\end{figure}

 In the following theorem we illustrate some fundamental properties of the proposed lifted operator. 
\begin{theorem}  \label{thm:lift_operator}
    The lifted operator $\mathcal{F}$ in Definition~\ref{defi:lift_operator} is causal. 
    Moreover, the lifted operator is passive from $u$ to $y$ if $\mathcal{G}$ is passive. That is, 
    \begin{equation}
    \langle\mathcal{P}_{\tau}u, \mathcal{P}_{\tau}\mathcal{F}(u,q)\rangle \geq 0
    \end{equation}
for all $(u,q)\in \ell_{2}\times\ell_2^{m_q}$ and all $\tau \in \mathbb{Z} $.    
\end{theorem}
\begin{proof}
    Causality of the operator follows directly from the assumptions on $\mathcal{N}$ and $\mathcal{G}$. 
    The claim of passivity is a direct extension of \cite[Prop. 2.11]{Rodolphe1996}. For instance,
    for any $(u,q) \in \ell_{2} \times \ell_{2}^{m_{q}}$, 
    take $y=\mathcal{N}(u,q)^{\dagger}\mathcal {G}\mathcal{N}(u,q)u$. Then, 
    $
    \langle u, y\rangle
    =
    \langle u, \mathcal{N}(u,q)^{\dagger}\mathcal {G}\mathcal{N}(u,q)u \rangle
    =
    \langle \mathcal{N}(u,q) u, \mathcal {G}\mathcal{N}(u,q)u \rangle
    \geq 0
    $, where the last inequality follows from the passivity of $\mathcal{G}$. Using causality, the general case follows directly from the last inequality: for all $(u,q) \in \ell_{2} \times \ell_{2}^{m_{q}}$
    and all $\tau \in \mathbb{Z}$,
    define $u_\tau \! = \!  \mathcal{P}_\tau u$.
    Then, 
    $\langle \mathcal{P}_\tau u, \mathcal{P}_\tau y\rangle  =  \langle u_\tau ,  \mathcal{P}_\tau\mathcal{N}(u_\tau,q)^{\dagger}\mathcal {G}\mathcal{N}(u_\tau,q)u_\tau\rangle = 
    \langle u_\tau, \mathcal{N}(u_\tau,q)^{\dagger}\mathcal {G}\mathcal{N}(u_\tau,q)u_\tau\rangle \geq 0
    $.  
\end{proof}
 
Theorem~\ref{thm:lift_operator} clarifies the main reason for the particular structure of Equation \eqref{eq:lift_operator} in Definition~\ref{defi:lift_operator}. 
The passivity of the linear filter $\mathcal{G}$ is a sufficient condition for the passivity of the nonlinear operator, for any given lifting operator $\mathcal{N}$. This allows us to borrow the passivity tools of LTI system theory to enforce passivity in the nonlinear setting.

\section{Passive lifted FIR operators} \label{sec:imple_operator}
We aim to use the lifted operator \eqref{eq:lift_operator} for parametric identification. Hence, in this section, we propose a finite-dimensional representation based on finite impulse response filters and parameterized nonlinear lifting filters. While alternative representations could be considered, this specific choice has the advantage of computational efficiency.

In what follows, the ($\theta_\mathcal{G}$-parameterized) linear operator $\mathcal{G}(\theta_{\mathcal{G}} ) \in 
\mathbb{B}(\ell_{2}^{m_{\mathcal{N}}},\ell_{2}^{m_{\mathcal{N}}})$ is the diagonal MIMO finite impulse response (FIR) filter
of the form
\begin{subequations}
\label{eq:G_FIR}
\begin{align} 
    \mathcal{G}(\theta_{\mathcal{G}} ) 
    &= 
    \begin{bmatrix}
        G(g_{0}) &                &        & 0 \\
                       & G(g_{1}) &        &   \\
                       &                & \ddots &   \\
        0              &                &        & G(g_{m_{\mathcal{N}}-1}) 
    \end{bmatrix}, \label{eq:Theta_mathcal_G}  \\
    \theta_\mathcal{G} 
     \in
    \Theta_{\mathcal{G}} 
    :\!\!&= 
    \{  (g_{0},\dots,g_{m_{\mathcal{N}}-1}) \,|\, g_{j}  \in \mathbb{R}^{m_{\mathcal{G}}}, 0 \leq \! j \! \leq m_{\mathcal{N}}-1  \}, \nonumber 
\end{align}
where $G(g_j)$ corresponds to the bi-infinite Toeplitz matrix representing the linear convolution associated with the $j$-filter finite impulse response $g_{j}$, $0 \leq \! j \! \leq m_{\mathcal{N}}\!-\!1$. Specifically, each $G(g_j)$ maps the input $u_j \in \ell_2$ to the output $y_j \in \ell_2$ of the SISO FIR filter associated to $g_j$,
    \begin{equation} \label{eq:fir_operator}
        y_j = G(g_j)u_j.
    \end{equation}
Note that $G(g)$ is causal and bounded by construction.
\end{subequations}

\begin{subequations} 
\label{eq:N_FIR}
For the parametrization of the lifting operator $\mathcal{N}$, we
rely on two key quantities: the vector of parameters
$\theta_{\mathcal{N}} \in \mathbb{R}^{m_{n}}$, the 
sample horizons $0< R_1 \in \mathbb{Z}$ and  $0< R_2 \in \mathbb{Z}$, and the family of functions
\begin{equation} \label{eq:f}
    f_{j}(\cdot,\cdot, \theta_{\mathcal{N}}) : \ell_{2[0,R_{1}-1]} \times \ell_{2[0,R_{2}-1]}^{m_{q}} \to \mathbb{R}
\end{equation}
for $0 \leq j \leq m_{\mathcal{N}}-1$. 
At implementation level, these functions will be 
typically defined by classical architectures for function approximation, such as neural networks, with `weights' $\theta_{\mathcal{N}}$. 
Each function
satisfies the following assumption 
\begin{assume}
\label{assume:bound_on_fj}
For all $0 \leq j \leq m_{\mathcal{N}}-1$,
each function $f_{j}(\cdot,\cdot, \theta_{\mathcal{N}})$ 
is continuous and 
there exists $c_j\geq 0$ such that 
\begin{equation}
|f_{j}(u,v,\theta_{\mathcal{N}})| \leq c_j
\end{equation}
for all $(u,v) \in  \ell_{2[0,R_{1}-1]} \times \ell_{2[0,R_{2}-1]}^{m_{q}}$. 
\end{assume}

Consider now the nonlinear operator 
$N(\cdot,\cdot, \theta_{\mathcal{N}}\!): \ell_{2} \!\times\! \ell_{2}^{m_{q}} \!\to\! \mathbb{B}(\ell_{2}^{m_{\mathcal{N}}},\ell_{2}^{m_{\mathcal{N}}})$
given by
    \begin{align} 
        &N(u,q,\theta_{\mathcal N})
:= \nonumber \\
&\begin{bmatrix}
N_0(u,q,\theta_{\mathcal N}) &  & & 0 \\
 & N_1(u,q,\theta_{\mathcal N}) &  &  \\
&  & \ddots &  \\
0 &  &  & N_{m_{\mathcal N}-1}(u,q,\theta_{\mathcal N})
\end{bmatrix},
    \end{align}
for all $(u,q) \in \ell_{2} \!\times\! \ell_{2}^{m_{q}}$,
where
each $N_{j}(u,q, \theta_{\mathcal{N}}) \in \mathbb{B}(\ell_{2},\ell_{2})$ reads
\begin{align}\label{eq:psi2}
    N_j(u,q,\theta_{\mathcal N})
&:=
\begin{bmatrix}
\ddots & \ddots &        &        &        \\
\ddots & n_{j,-1} & 0      &        &        \\
& 0      & n_{j,0}  & 0      &        \\
&        & 0      & n_{j,1}  & \ddots \\
&        &        & \ddots & \ddots 
\end{bmatrix} \nonumber \\
n_{j,t}
&:=
f_j(\bar{\mathcal P}^{R_1}_{t}u,\bar{\mathcal P}^{R_2}_{t}q, \theta_{\mathcal N}),
\end{align}
for all $u,q \in \ell_{2} \!\times\! \ell_{2}^{m_{q}}$
and all $0 \leq j \leq m_{\mathcal{N}-1}$.
Note that each $N_{j}(\cdot,\cdot, \theta_{\mathcal{N}})$ is a self-adjoint operator by construction $N_{j}(\cdot,\cdot, \theta_{\mathcal{N}}) = N_{j}(\cdot,\cdot, \theta_{\mathcal{N}})^{\dagger}$. Likewise,
\begin{equation}
N(\cdot,\cdot, \theta_{\mathcal{N}}) = N(\cdot,\cdot, \theta_{\mathcal{N}})^{\dagger}.
\end{equation}

The diagonal structure of $N$ mirrors the 
one of $\mathcal{G}(\theta_{\mathcal{G}})$ and it is instrumental
to establish an `element-wise' nonlinear lifting for each
FIR filter. With these quantities, the
lifting operator $\mathcal{N}(\cdot,\cdot, \theta_{\mathcal{N}})$ is defined as
\begin{align}
    \mathcal{N}(u,q, \theta_{\mathcal{N}})u := N(u,q, \theta_{\mathcal{N}}) \mathbf{1}_{m_{\mathcal{N}}}u \label{eq:N_cal_defi}
\end{align}
for all $u \in \ell_{2}$ and $q\in \ell_2^{m_q}$, 
where $\mathbf{1}_{m_{\mathcal{N}}} \in 
\mathbb{B}(\ell_{2},\ell_{2}^{m_{\mathcal{N}}})$ satisfies
\begin{align}
    \mathbf{1}_{m_{\mathcal{N}}} u: = 
    \begin{bmatrix}
        u \\
        \vdots \\
        u
    \end{bmatrix}.
\end{align}
for all $u \in \ell_{2}$. Likewise,
$\mathcal{N}(\cdot,\cdot, \theta_{\mathcal{N}})^{\dagger}$ is defined as
\begin{align}
    \mathcal{N}(u,q, \theta_{\mathcal{N}})^{\dagger}z := \mathbf{1}_{m_{\mathcal{N}}}^{\dagger}  N(u,q, \theta_{\mathcal{N}}) z \label{eq:N_adj_cal_defi}
\end{align}
for all $u \in \ell_{2}$, $q\in \ell_2^{m_q}$ and $z\in \ell_2^{m_\mathcal{N}}$, where
the adjoint  $\mathbf{1}_{m_{\mathcal{N}}}^{\dagger}\in \mathbb{B}(\ell_{2}^{m_{\mathcal{N}}},\ell_{2})$ satisfies 
\begin{align}
    \mathbf{1}_{m_{\mathcal{N}}}^{\dagger} \begin{bmatrix}
        z_{0} \\
        \vdots \\
        z_{m_{\mathcal{N}-1}}
    \end{bmatrix} &:= \sum_{j=0}^{m_{\mathcal{N}-1}}z_{j}
\end{align}
for all $z_0, \dots, z_{m_{\mathcal{N}-1}} \in \ell_2$.
\end{subequations}

We can thus write the lifted operator \eqref{eq:lift_operator} as 
\begin{align} 
    y &=\mathcal{N}(u,q,\theta_{\mathcal{N}})^{\dagger}\mathcal{G}(\theta_{\mathcal{G}})\mathcal{N}(u,q,\theta_{\mathcal{N}})u \nonumber \\
    &= \mathbf{1}_{m_{\mathcal{N}}}^{\dagger}  N(u,q, \theta_{\mathcal{N}})^{\dagger}\mathcal{G}(\theta_{\mathcal{G}})N(u,q, \theta_{\mathcal{N}})\mathbf{1}_{m_{\mathcal{N}}}u \label{eq:NFIR_operator2} \\
    & = \sum_{j=0}^{m_{\mathcal{N}}-1}N_{j} (u,q,\theta_{\mathcal{N}})  G (g_{j}) N_{j} (u,q,\theta_{\mathcal{N}}) u. \nonumber
\end{align}  
We denote \eqref{eq:NFIR_operator2} as a nonlinear FIR (NFIR) operator since it is a (finitely represented) nonlinear lifting of a linear FIR filter $\mathcal{G}$. To compute the current sample of the output $y$, the filter requires $\max(m_\mathcal{G},R_1)$ past samples of the input $u$ and $R_2$ past samples of the additional input $q$. 

\begin{definition}{NFIR operator} \label{defi:NFIR2}
    For a given \emph{operator size} $(m_{\mathcal{N}},R_1,R_2,m_{\mathcal{G}},m_n)\in\mathbb{N}^5$
    and \emph{operator parameters} 
    \textcolor{black}{$\theta_{\mathcal{G}} \in \mathbb{R}^{ m_{\mathcal{N}} m_\mathcal{G}}$}
    and $\theta_{\mathcal{N}} \in \mathbb{R}^{m_{n}}$,
    the \emph{nonlinear finite-impulse response} 
    operator $\mathcal{F}_{\mathrm{NFIR}}(\cdot,\cdot,\theta_{\mathcal{G}},\theta_{\mathcal{N}}): \ell_2\times\ell_2^{m_q} \to \ell_2$
    \begin{equation}
    y = \mathcal{F}_{\mathrm{NFIR}}(u,q,\theta_{\mathcal{G}},\theta_{\mathcal{N}}) \label{eq:NFIR_operator}
    \end{equation}
    satisfies \eqref{eq:G_FIR},\eqref{eq:N_FIR},\eqref{eq:NFIR_operator2}
    for all inputs $(u,q) \in \ell_{2}\times \ell_{2}^{m_{q}}$ and all
    outputs $y \in \ell_{2}$.
\end{definition}

The following theorem shows that the NFIR operators retain the key properties of 
lifted passive operators in Theorem~\ref{thm:lift_operator}. 
\begin{theorem} \label{thm:NFIR_passive}
For a given size $(m_{\mathcal{N}},R_1,R_2,m_{\mathcal{G}},m_n)\in\mathbb{N}^5$ and parameters
    $\theta_{\mathcal{G}} \in \mathbb{R}^{ m_\mathcal{N}  m_\mathcal{G}}$
    and $\theta_{\mathcal{N}} \in \mathbb{R}^{m_{n}}$,
    the NFIR 
    operator $\mathcal{F}_{\mathrm{NFIR}}(\cdot,\cdot,\theta_{\mathcal{G}},\theta_{\mathcal{N}})$ of Definition \ref{defi:NFIR2}
    is causal and has finite $\ell_2$ gain.     Moreover, the lifted operator is passive from $u$ to $y$ if $\mathcal{G}$ is passive. That is, 
    \begin{equation}
    \langle\mathcal{P}_{\tau}u, \mathcal{P}_{\tau}\mathcal{F}_{\mathrm{NFIR}}(u,q,\theta_{\mathcal{G}},\theta_{\mathcal{N}})\rangle \geq 0
    \end{equation}
for all $(u,q)\in \ell_{2}\times\ell_2^{m_q}$ and all $\tau \in \mathbb{Z} $.
\end{theorem}
\begin{proof}
    Causality follows from the fact that all operators in \eqref{eq:NFIR_operator2} are causal. For the $\ell_2$ gain we have    \begin{align}
    &\Vert \mathcal{F}_{\mathrm{NFIR}}(u,q,\theta_{\mathcal{G}},\theta_{\mathcal{N}}) \Vert_{2} \nonumber \\
    &\leq 
    \sum_{j=0}^{m_{\mathcal{N}}-1} 
    \Vert N_{j} (u,q,\theta_{\mathcal{N}})\Vert_{2}  \Vert G (g_{j})\Vert_{2} \Vert N_{j} (u,q,\theta_{\mathcal{N}})\Vert_{2} \Vert u \Vert_{2} \nonumber \\
    &\leq 
     \sum_{j=0}^{m_{\mathcal{N}}-1}
      {c_j}^2 \Vert G (g_{j})\Vert_{2} 
      \Vert u \Vert_{2}
    \leq 
    \left( \sum_{j=0}^{m_{\mathcal{N}}-1}
      c_j^2 \gamma_j\right)
      \Vert u \Vert_{2}
    \end{align}
    where $c_j$ is from Assumption \ref{assume:bound_on_fj} and $\gamma_j$ is the gain of each linear filter 
    $G(g_j)$.
    Finally, given the particular structure of 
    \eqref{eq:NFIR_operator2}, the proof of the operator passivity follows the same steps of the proof of Theorem \ref{thm:lift_operator}. 
\end{proof}
    
Equation~\eqref{eq:NFIR_operator2} can be interpreted as 
the
parallel connection of $m_{\mathcal{N}}$ branches $N_{j}(u,q,\theta_{\mathcal{N}})^{\dagger}G(g_{j}) N_{j}(u,q,\theta_{\mathcal{N}}) u$. 
Fig.~\ref{fig:nfir_block_diagram} provides a block diagram of branch $j$, highlighting the case where the nonlinear lifting is made recursive via the feedback $q(t) = y(t-1)$ for all $t \in \mathbb{Z}$. 
\begin{figure}[htbp]
    \centering
\begin{tikzpicture}[
    scale=0.65,
    transform shape,
    >=Stealth,
    signal/.style={,-{Stealth[length=2.6mm,width=2mm]}},
    thicksignal/.style={,-{Stealth[length=2.8mm,width=2.2mm]}},
    wire/.style={},
    block/.style={draw, , minimum height=1.15cm, minimum width=1.9cm, align=center},
    smallblock/.style={draw, , minimum height=1.02cm, minimum width=1cm, align=center},
    circblock/.style={draw, circle, , minimum size=0.95cm, inner sep=0pt}
]

    \node[circblock] (Pi1) at (6.9,2.85) {$\Pi$};
    \node[block, minimum width=2.1cm, minimum height=1.05cm] (Gamma) at (9.25,2.85)
        {$G(g_j)$};
    \node[circblock] (Pi2) at (11.75,2.85) {$\Pi$};

    \node[block, minimum width=2.35cm, minimum height=1.10cm] (psiblock) at (4.25,1.45)
        {$f_{j}(\bar{\mathcal{P}}_{t}^{R_1} u,\bar{\mathcal{P}}_{t}^{R_2} q, \theta_{\mathcal{N}}) $};

    \node[smallblock] (delay) at (6.95,0.20)
        {$z^{-1}$};

    \draw[signal] (0.35,2.85) -- (Pi1.west);
    \draw[signal] (Pi1.east) -- (Gamma.west);
    \draw[signal] (Gamma.east) -- (Pi2.west);
    \draw[signal] (Pi2.east) -- (13.70,2.85);

    \node[above=2pt] at (0.95,2.85) {$u(t)$};
    \node[above=2pt] at (12.95,2.85) {$y(t)$};

    \draw[wire] (1.55,2.85) -- (1.55,1.65);
    \draw[signal] (1.53,1.65) -- (2.55,1.65) ;

    \draw[wire] (13.10,2.85) -- (13.10,0.20);
    \draw[wire] (13.10,0.20) -- (delay.east);
    \draw[wire] (delay.west) -- (1.53,0.20);
    \draw[wire] (1.55,0.20) -- (1.55,1.15);
    \draw[signal] (1.53,1.15) -- (2.55,1.15);

    \draw[wire] (psiblock.east) -- (11.75,1.45);

    \draw[signal] (6.90,1.45) -- (Pi1.south);
    \draw[signal] (11.75,1.45) -- (Pi2.south);

\end{tikzpicture}

    \caption{Block diagram of the operator. $z^{-1}$ is the delay operator.}
    \label{fig:nfir_block_diagram}
\end{figure}

\section{Learning passive lifted FIR operators}\label{sec:freq_vs_time}
Given a set $\mathcal{S}$ of $m_\mathcal{S} \in \mathbb{N}$ measured input-output data of finite length\footnote{For notational simplicity we consider these as signals in $\ell_{2}$ and $\ell_{2}^{m_{q}}$.}
\begin{align}
\mathcal{S} &:= \{(u_{i},q_{i},y_{i}) :
0 \leq  i \leq {m_{\mathcal{S}} \!-\! 1}\} \subseteq \ell_{2} \times \ell_{2}^{m_{q}} \times \ell_{2}, 
\end{align}
the (regularized) NFIR operator $\mathcal{F}_{\mathrm{NFIR}}$ that best fits $\mathcal{S}$ satisfies the following optimization problem: given the 
operator size $(m_{\mathcal{N}},R_1,R_2,m_{\mathcal{G}},m_n)\in\mathbb{N}^5$
and weights $ 0 \leq \gamma_\mathcal{G}, \gamma_\mathcal{N} \in \mathbb{R}$,
\begin{subequations} \label{eq:rand1}
    \begin{equation} 
       \underset{\theta_{\mathcal{G}} \in \Theta_{\mathcal{G}},\theta_{\mathcal{N}} \in \mathbb{R}^{m_{n}} }{\min}  J_\mathcal{S}(\theta_{\mathcal{G}},\theta_{\mathcal{N}}) 
       + \gamma_\mathcal{G} \lVert \theta_\mathcal{G}\rVert_{2}^{2}
       + \gamma_\mathcal{N} \lVert \theta_\mathcal{N}\rVert_{2}^{2}
       \label{eq:data_set}
       \end{equation}
       where
       \begin{align} 
       &J_\mathcal{S}(\theta_{\mathcal{G}},\theta_{\mathcal{N}})  \label{eq:J_Theta_G_Theta_L} \\
       &=\sum_{i=0}^{ m_{\mathcal{S}} -1} 
       \Big\lVert y_{i} \!-\! \sum_{j=0}^{m_{\mathcal{N}}-1}N_{j}(u_{i},q_{i},\theta_{\mathcal{N}})G(g_{j}) N_{j}(u_{i},q_{i},\theta_{\mathcal{N}}) u_{i} \Big\rVert^2_2. \nonumber 
       \end{align}  
\end{subequations}
 $ J_\mathcal{S}(\theta_{\mathcal{G}},\theta_{\mathcal{N}})$ is \emph{not convex} when jointly optimizing $\theta_{\mathcal{G}}$ and $\theta_{\mathcal{N}}$. However, for any given $\theta_{\mathcal{N}}$, the identification 
of the optimal linear FIR filter, 
    \begin{equation} 
       \underset{\theta_{\mathcal{G}} \in \Theta_{\mathcal{G}}}{\min}  J_\mathcal{S}(\theta_{\mathcal{G}},\theta_{\mathcal{N}})
       + \gamma_\mathcal{G} \lVert \theta_\mathcal{G}\rVert_{2}^{2},
       \label{eq:convex_linear}
       \end{equation}
reduces to a standard \emph{convex least-squares problem}.
Learning 
an 
NFIR filter is thus easy when the lifting is
 pre-determined, i.e., when the family of functions $f_j$ in \eqref{eq:f} and the weights $\theta_{\mathcal{N}}$ are given. This suggests that the co-optimization of lifting and linear FIR filters can be approached iteratively, through alternating optimization.

For passivity, we can take advantage of Theorem \ref{thm:NFIR_passive}. We just need to constrain the linear
 part of the NFIR filter to be passive, 
 by adapting \cite[Thm. 4]{Wang2024}.
 In the following, we write $g(k)$ to
denote the $k$-th sample of the impulse response $g$.
\begin{theorem} \label{thm:passivity_G}
Consider any parameter vector $\theta_{\mathcal{G}} = (g_{0},\dots,g_{m_{\mathcal{N}}-1})$, where $g_{j}  \in \mathbb{R}^{m_{\mathcal{G}}}$ for $j \in \{0, \dots,  m_{\mathcal{N}}-1\}$,
any \emph{sampling horizon}
$H \in \mathbb{Z}_{> 0}$, any \emph{bound} $\rho \in \mathbb{R}$, $\rho >0 $, and \emph{decay rates}
$\rho_j \in \mathbb{R}$, $ 0 < \rho_j < 1 $ for $j \in \{0, \dots, m_{\mathcal{N}} -1\}$.

For $j \in \{0,\dots,m_{\mathcal{N}}-1\}$,
there exist $\epsilon_{j} \geq 0$ such that, if 
    \begin{subequations} \label{eq:freq_passivity}
        \begin{align} 
                 \sum_{k=0}^{m_{\mathcal{G}}-1} g_j(k) \left( e^{ -\frac{\textcolor{black}{\mathrm{i}}  h \pi}{H} k} + e^{ \frac{\textcolor{black}{\mathrm{i}}  h \pi}{H} k} \right) \geq \epsilon_{j} \label{eq:frequency^constraint}\\
                 - \rho\rho_{j}^{k} \leq g_{j}(k) \leq \rho\rho_{j}^{k} \label{eq:infirsSamp3} 
                 \end{align}  
                 for all $k \in \{ 0,\dots, m_{\mathcal{G}} - 1\}$,
                 $j \in \{0, \dots,  m_{\mathcal{N}}-1\}$
                 and $ h\in\{0,\dots,H\}$,  
    \end{subequations}
    then $\mathcal{G}(\theta_{\mathcal{G}} )$ in \eqref{eq:G_FIR} is passive.
\end{theorem}

Remarkably, \eqref{eq:convex_linear} and \eqref{eq:freq_passivity} define a constrained convex optimization problem. Furthermore, by Theorem \ref{thm:NFIR_passive}, the combination of \eqref{eq:rand1} and \eqref{eq:freq_passivity} leads to passive 
NFIR filters that best fit the data.
\textcolor{black}{Equation}~\eqref{eq:frequency^constraint} is simply a sampled positive-real condition on the 
transfer function of the FIR filters
within $\mathcal{G}(\theta_\mathcal{G})$.
The role of $\epsilon_j$ is to guarantee
that positive realness is not violated at inter-sample
frequencies. A conservative bound on $\epsilon_j$ can be found in
\cite{{Wang2024}}: 
\begin{equation}
\label{eq:frequency_bound}
\epsilon_{j} \geq \pi\rho \frac{1-\rho_{j}^{m_{\mathcal{G}}}}{1-\rho_{j}} \frac{m_{\mathcal{G}}-1}{2H}, \
\end{equation}
which gets smaller as the sampling horizon $H$ increases,
and/or the bound $\rho$ and the decay rate $\rho_j$ decreases. However, in practice, we will see that it is much more convenient to  \emph{guess} reasonably small $\epsilon_j$, then check the passivity of the FIR filter $\mathcal{G}(\theta_\mathcal{G})$ post optimization.
Equation~\eqref{eq:infirsSamp3} enforces
the exponential decay of the tail of the impulse response of each FIR filter. When combined to \eqref{eq:convex_linear}, this form of regularization 
can be used to remove undesired high-frequency components in the response of the FIR filter\footnote{We defer a proper discussion on parameter tuning to the next section.}.

Another passivity constraint is based on a time-domain formulation (KYP lemma) as shown in \cite[Thm. 1]{Wang2024},
revisited here for completeness.
\begin{theorem}
\label{thm:KYP}
Consider any parameter vector $\theta_{\mathcal{G}} = (g_{0},\dots,g_{m_{\mathcal{N}}-1})$, where $g_{j}  \in \mathbb{R}^{m_{\mathcal{G}}}$ for $j \in \{0, \dots,  m_{\mathcal{N}}-1\}$. Define the state-space matrices
\begin{subequations}
\label{eq:ss_passivity}
    \begin{align}
        A &= \begin{bmatrix}
            0 & 1 & 0 &0 &\dots &0 \\
            0 & 0 & 1 &0 &\dots &0 \\
            \vdots & & & & & \\
            0 & 0 & 0 &0 &\dots &1 \\
            0 & 0 & 0 &\dots & &0 
        \end{bmatrix} , \quad  
        B = \begin{bmatrix}
            0\\
            0  \\
            \vdots \\
            0 \\
            1
        \end{bmatrix}  \\
        \textcolor{black}{C_{j}} &= 
        \begin{bmatrix}
            g_{j}(m_{\mathcal{G}} - 1) & \dots & g_{j}(2) & g_{j}(1)
        \end{bmatrix}, \quad 
        D_{j}= g_{j}(0) \, . \nonumber 
    \end{align}
    for $j \in \{0, \dots,  m_{\mathcal{N}}-1\}$.
    If the following LMIs in the unknown matrices 
    $ X_{j} \in \mathbb{R}^{(m_{\mathcal{G}} - 1)\times (m_{\mathcal{G}} - 1)} $ are feasible for all
    $j \in \{0, \dots,  m_{\mathcal{N}}-1\}$
      \label{eq:KYP_LMIs}
        \begin{align}
                X_{j}=X_{j}^T & >  0 \nonumber \\
                \quad \begin{bmatrix}
                X_{j} - A^{T}X_{j}A & C_{j}^{T} - A^{T}X_{j}B \\
                C_{j}-B^{T}X_{j}A  & D_{j} + D_{j}^{T} - B^{T}X_{j}B
                \end{bmatrix} &\geq 0 \, \\
                D_{j} + D_{j}^{T} &\geq 0, \nonumber  
        \end{align}
    \end{subequations}
    then $\mathcal{G}(\theta_{\mathcal{G}} )$ in \eqref{eq:G_FIR} is passive.
\end{theorem}

As above, the combination of \eqref{eq:convex_linear} and 
\eqref{eq:ss_passivity}  gives a constrained convex optimization problem. Likewise,  \eqref{eq:rand1},
\eqref{eq:ss_passivity}, and 
Theorem \ref{thm:NFIR_passive}
provide another route to derive passive NFIR filters from data.
In contrast to \eqref{eq:freq_passivity},
\eqref{eq:KYP_LMIs} does not require any parameter tuning.
At the same time, when the FIR order increases, LMI  constraints may not be satisfied as efficiently as \eqref{eq:freq_passivity}. 
This is illustrated in Section~\ref{sec:examples}.

We can finally address the co-learning of  lifting function parameters $\theta_{\mathcal{N}}$ and FIR filters parameters $\theta_{\mathcal{G}}$. Since \eqref{eq:rand1} is non-convex, we propose the
iterative approach of Algorithm~\ref{alg:alternating}.
Each iteration of Algorithm~\ref{alg:alternating} updates first $\theta_{\mathcal{G}}$ then $\theta_{\mathcal{N}}$ . The update of $\theta_{\mathcal{G}}$ is derived via convex constrained optimization,
combining \eqref{eq:convex_linear} with
either \eqref{eq:freq_passivity}
or \eqref{eq:ss_passivity}, as discussed above. For the first update, we randomly initialize $\theta_{\mathcal{N}}$.
We use CVXPY \cite{bib:cvxpy} with MOSEK \cite{bib:mosek} solver for this step.  
The next part of the iteration freezes $\theta_{\mathcal{G}}$ to update $\theta_{\mathcal{N}}$. Here, the convexity of the optimization problem
fundamentally depends on the structure of the lifting function. Typically, this is not a convex optimization, which we approach using ADAM solver with an analytically computed gradient. 
The algorithm alternates between the two updates until the maximum iteration number is reached. The effectiveness of Algorithm~\ref{alg:alternating} is illustrated in the examples of Section~\ref{sec:examples}. 

\begin{algorithm}[htbp]
\label{alg:main}
\caption{Alternating algorithm for \eqref{eq:rand1}}
\label{alg:alternating}
    \begin{algorithmic}[1]
        \REQUIRE NFIR operator (Definition \ref{defi:NFIR2}), input-output data 
        $\mathcal S=\{(u_i,q_i,y_i)\}_{i=0}^{m_{\mathcal{S}}-1}$,
        cost function \eqref{eq:rand1},
        initial guess $\theta_{\mathcal N}^{(-1)}\in \mathbb{R}^{m_{n}}$,
        and maximum number of iterations $K\geq 1$.
        \vspace{0mm}
        
        \FOR{$k=0,1,\ldots,K-1$}
        
        \STATE \textbf{$\theta_{\mathcal G}$ update:} 
        \begin{subequations}
       \begin{equation}
        \theta_{\mathcal G}^{(k)}
        =
        \arg\min_{\theta_{\mathcal G}\in\Theta_{\mathcal G}}
        J_\mathcal{S}\!\left(\theta_{\mathcal G},{\theta}_{\mathcal N}^{(k-1)}\right)
        +
        \gamma_\mathcal{G}
        \|\theta_{\mathcal G}\|_2^2, \label{eq:alg_FIR}
        \end{equation}
        subject to either \eqref{eq:freq_passivity} or \eqref{eq:KYP_LMIs}.
        
        \STATE \textbf{$\theta_{\mathcal N}$ update:} 
        \begin{equation}
            \theta_{\mathcal N}^{(k)}
            =
            \arg\min_{\theta_{\mathcal N}\in \mathbb{R}^{m_{n}}}
            J_\mathcal{S}\!\left(\theta_{\mathcal G}^{(k)},\theta_{\mathcal N}\right)        +
        \gamma_\mathcal{N}
        \|\theta_{\mathcal N}\|_2^2 
            . \label{eq:alg_LIFT}
        \end{equation}
        \end{subequations}
        \ENDFOR
        
        \STATE \RETURN $\theta_{\mathcal G}^{(K-1)},\theta_{\mathcal N}^{(K-1)}$.
    \end{algorithmic}
\end{algorithm}

\section{Hyper-parameters and lifting functions}\label{sec:hyper_lift}

\subsection{Tuning of hyper-parameters.}\label{sec:hyperpara_tune}

Here we provide insights on the selection of $m_{\mathcal{N}}$, $R_1$, $R_2$, $m_{\mathcal{G}}, H, \rho, \gamma_{\mathcal{G}}, \gamma_{\mathcal{N}}$ and $\rho_{j},  \epsilon_{j}$ for $j \in \{0,\dots,m_{\mathcal{N}}-1\}$. We start by focusing on
their role in shaping the features of the NFIR operator. 
\emph{Passivity:} $H$, $\epsilon_{j}$, $\rho$, and $\rho_j$ control passivity. As shown in \eqref{eq:frequency_bound}, smaller frequency sampling $H$ reduces the number of constraints (lower computational effort) but requires a more conservative bound $\epsilon_j$. Similar observations hold for $\rho$ and $\rho_j$.
\emph{Gain:} $\rho$ regulates the gain of the FIR filters; therefore,
indirectly, it affects the gain of the NFIR operator. A larger (smaller) $\rho$ enables a larger (smaller) gain.
\emph{Time scale matching:} 
The decay rate $\rho_j$ and
the quantity $m_{\mathcal{G}}T_{s}$, where $T_{s}$ is the sampling time,  should be tuned to match the time scale of the (fading-memory) operator producing the data. 
For example, 
\textcolor{black}{if}
an experimental step response comes within $10 \%$ of its steady-state value at $t=2$ seconds, then $m_{\mathcal{G}}T_{s} > 2$ and $\rho_j$ should be tuned accordingly. Here, we are assuming that the time scale of the NFIR is dominated by the time scale of the FIR filters, which is  reasonable when the memory of the lifting operator $R_{1}, R_{2} \ll m_{\mathcal{G}}$ in~\eqref{eq:f}. 
\emph{Complexity and expressivity:} 
The parameter $m_{\mathcal{N}}$ regulates the complexity of the lifting operator and, consequently, the expressivity of the NFIR operator. This is because a larger $m_{\mathcal{N}}$ enforces a lifting into a higher-dimensional space, whereas a smaller $m_{\mathcal{N}}$ restricts it to a lower-dimensional one.

These parameters also shape the optimization problem \eqref{eq:rand1}.
\emph{Regularization:}
The constants $\gamma_{\mathcal{G}}, \gamma_{\mathcal{N}}$ should be increased to reduce overfitting. Even with negligible noise, small non-zero regularization parameters are needed to guarantee that the least-squares problem in \eqref{eq:convex_linear} is not ill-conditioned. 
\emph{Feasibility and uniqueness of the solution:}
    Equation~\eqref{eq:convex_linear} is a strictly convex $\ell_{2}$-regularized least-squares problem. A unique solution exists provided that $\gamma_{\mathcal{G}} > 0$ and the passivity constraints are feasible. 
    For the latter, the frequency-domain constraints \eqref{eq:freq_passivity} are 
    always feasible if $\epsilon_{j} \leq 2|g_{j}(0)|$
    and $\rho > 0$. In fact, under these conditions, taking
    $0 < g_{j}(0) \leq \rho$ and $g_{j}(k)=0$ for all $k>0$ and all   $0 \leq j \leq m_{\mathcal{N}}-1$ 
    gives a passive FIR filter $\mathcal{G}(\theta_\mathcal{G})$.
    Also the time-domain constraints \eqref{eq:ss_passivity} are always feasible, since a solution exists if and only if the matrices $A$, $B$, $C_{j}$, and $D_{j}$ are the state-space realization of a passive filter. A trivial case is thus given by 
    $g_{j}(0)>0 $ and $g_{j}(k)=0$ for all $k>0$ and all   $0 \leq j \leq m_{\mathcal{N}}-1$.

\subsection{Choice of the lifting functions}\label{sec:choice_lifting}

A theoretical characterization of the expressivity of NFIR operators is challenging and inherently related to the choice of the family of lifting functions \eqref{eq:f}.  However, \eqref{eq:N_FIR} allows for the adoption of general-purpose lifting functions, such as neural networks, and their efficient training algorithms (including regularization approaches).
Many architectures could be considered, as long as Assumption~\ref{assume:bound_on_fj} is 
\textcolor{black}{satisfied.}

As an example, consider the multilayer perceptron (MLP) below. The network has  three layers of dimensions $m_{W_{1}}$, $m_{W_{2}}$, and 
$m_{\mathcal{N}}$
respectively, and  activation function $\sigma$. The examples of 
Section~\ref{sec:examples}
will adopt $\sigma = \tanh$. 
\begin{subequations}
\label{eq:NN_for_lifting}
Define

\begin{equation}
p(t) =
\begin{bmatrix}
        p_u(t) \\
        p_q(t)\\
    \end{bmatrix} =
\begin{bmatrix}
        \bar{\mathcal{P}}_{t}^{R_1} u \\
        \bar{\mathcal{P}}_{t}^{R_2} q\\
    \end{bmatrix} \in \mathbb{R}^{R_{1}+m_{q}R_{2}},
\end{equation}
and take
\begin{align}
    &
    \begin{bmatrix}
        f_{0}(p_u(t),p_q(t),\theta_{\mathcal{N}}) \\
        \vdots \\
         f_{m_{\mathcal{N}-1}}(p_u(t),p_q(t),\theta_{\mathcal{N}}) 
    \end{bmatrix}  \nonumber \\
    &= \sigma(W_{3}\sigma(W_{2}\sigma(W_{1}p(t) + b_{1})+b_{2}) + b_{3}) \in \mathbb{R}^{m_{\mathcal{N}}}, \label{eq:f_as_NN}
\end{align}
where 
\begin{align}
    &W_{1} \in \mathbb{R}^{m_{W_{1}} \times (R_{1}+m_{q}R_{2})}, \ b_{1} \in \mathbb{R}^{m_{W_{1}}} \nonumber \\
    &W_{2} \in \mathbb{R}^{m_{W_{2}} \times m_{W_{1}} }, \ b_{2} \in \mathbb{R}^{m_{W_{2}}} \\
    &W_{3} \in \mathbb{R}^{m_{\mathcal{N}} \times m_{W_{2}} }, \ b_{3} \in \mathbb{R}^{m_{\mathcal{N}} } \nonumber
\end{align} 
and $\theta_{\mathcal{N}} $ 
is formed by stacking the vectorized matrices $W_1$, $W_2$, $W_3$, and the 
vectors
$b_1$, $b_2$, $b_3$.
\end{subequations}

\subsection{Feedback and training }\label{sec:simulation_error}
The NFIR operator of Definition~\ref{defi:NFIR2} becomes recursive when the input $q$ depends on the operator's output $y$. Specifically, we enforce strict causality by setting $q(k)=y(k - \tau)$ for all time steps $k$, where $\tau > 0$. For recursive NFIR operators there is a crucial difference between training and deployment. After training, the output of the NFIR operator is fed back to its input to generate the next output sample (closed-loop). In contrast, during training based on Algorithm~\ref{alg:alternating}, $q$ is typically supplied by the data, for computational efficiency (open-loop). In the latter case, Algorithm~\ref{alg:alternating} minimizes the one-step-ahead prediction error in the sense of~\cite[Chp.~3]{Ljung_sysidBook_1999}.

To minimize the simulation error during the deployment phase  due to the recursive generation of  $q$ ~\cite[Chp. 3]{Ljung_sysidBook_1999}, 
the update of $\theta_{\mathcal{N}}$ in Algorithm~\ref{alg:alternating}
can be replaced by backpropagation through time,
at the cost of 
\textcolor{black}{increasing}
computational complexity.
We illustrate the effect of this modification in  Section~\ref{sec:examples} where
we keep Algorithm~\ref{alg:alternating} as it is but replace the last iteration over $\theta_{\mathcal{N}}$ with one instance of backpropagation. The modification of the last step increases the quality of
the estimation while mitigating computational effort.

\section{Examples: identification of passive systems} \label{sec:examples}

\subsection{Nonlinear mass-spring-damper system (sim data)}
\label{example:mass-spring}

\begin{figure}[htbp]
  \centering
  \resizebox{0.45\columnwidth}{!}{%
      \begin{tikzpicture}[
            every node/.style={outer sep=0pt},
            thick,
            mass/.style={draw,thick},
            spring/.style={
                thick,
                decorate,
                decoration={
                    zigzag,
                    pre length=0.3cm,
                    post length=0.3cm,
                    segment length=6
                }
            },
            ground/.style={
                fill,
                pattern=north east lines,
                draw=none,
                minimum width=0.75cm,
                minimum height=0.3cm
            },
            dampic/.pic={
                \fill[white] (-0.1,-0.3) rectangle (0.3,0.3);
                \draw (-0.3,0.3) -| (0.3,-0.3) -- (-0.3,-0.3);
                \draw[line width=1mm] (-0.1,-0.3) -- (-0.1,0.3);
            }
        ]
        
            \node[
                mass,
                minimum width=3.5cm,
                minimum height=2cm,
                fill=none
            ] (m1) {$m$};
        
            \node[
                left=2cm of m1,
                ground,
                minimum width=3mm,
                minimum height=2.5cm
            ] (g1) {};
            \draw (g1.north east) -- (g1.south east);
        
            \draw[spring] ([yshift=3mm]g1.east) coordinate(aux)
                -- (m1.west|-aux)
                node[midway,above=1mm] {$k$};
        
            \draw ([yshift=-3mm]g1.east) coordinate(aux')
                -- (m1.west|-aux')
                pic[midway] {dampic}
                node[midway,below=3mm] {$c$};
        
            \draw[thin] (m1.north) -- ++ (0,1) coordinate[midway](aux1);
            \draw[latex-] (aux1) -- ++ (-0.5,0) node[above] {$F(t)$};
        
            \draw[thin,dashed] (m1.south) -- ++ (0,-1) coordinate[pos=0.85](aux'1);
            \draw[latex-] (aux'1) -- ++ (-1,0)
                node[midway,above] {$ \quad \dot x(t)$}
                node[left,ground,minimum height=7mm,minimum width=1mm] (g'1) {};
            \draw[thick] (g'1.north east) -- (g'1.south east);
        \end{tikzpicture}%
    }
    \includegraphics[width=0.49\columnwidth]{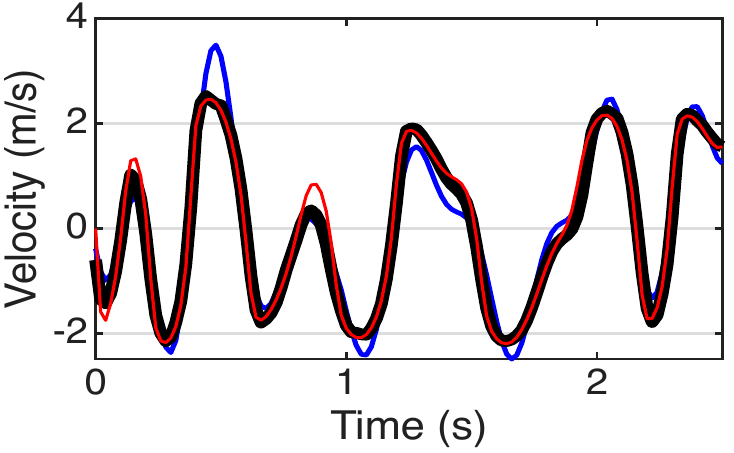} 
  \caption{Left: a mass-spring-damper system with mass $m=\SI{0.1}{\kilogram}$, linear spring stiffness \(k=\SI{10}{\newton\per\metre}\). The damper is nonlinear. $u(t) = F(t)$ is the input force. $ y(t) = \dot x(t)$ is the velocity output. Right: responses for models identified from noisy output measurements. Red: noise-free version of output $y$. Black: NFIR output. Blue: FIR output.} 
  \label{fig:mass_spring_damper}
\end{figure}

We consider the mass-spring-damper system in Fig.~\ref{fig:mass_spring_damper}, where 
the force exerted by the nonlinear damper $c$ satisfies
\[
    f_{c} = - 0.5\dot x - {\dot x}^{3}.
\]
We consider a sampling time of \(\SI{0.02}{\second}\) and perturb the system input with a forcing signal given by the summation of 10 random-phase sine waves. 
The forcing signal lasts \(\SI{5}{\second}\), 
\textcolor{black}{with frequencies}
being linearly spaced in the range \([0,7.5 \pi]~\si{\radian\per\second}\),
the amplitude of each sine wave being $15$ N. 

As first step, we use Algorithm \ref{alg:alternating}
to complete a linear system identification based on
a single FIR operator (no nonlinear lifting). This removes \eqref{eq:alg_LIFT}
from the optimization. We run the algorithm for different FIR orders, considering both frequency-domain
and time-domain  constraints \eqref{eq:freq_passivity}
and \eqref{eq:KYP_LMIs}. 
Simulation results are provided in Fig.~\ref{fig:mass_spring_damper} ($m_\mathcal{G} = 50$, in blue). Computing times  for different FIR orders $m_{\mathcal{G}}$ are summarized in Table~\ref{tab:time},
based on an Apple M2 processor,  It can be seen that the frequency-domain constraints in  Theorem~\ref{thm:passivity_G} are faster and scale better for increasing FIR order.

\begin{table}[htbp]
\centering
\caption{Computation time for FIR: LMIs vs frequency constraints.}
\label{tab:time}
\scriptsize
\setlength{\tabcolsep}{3pt}
\renewcommand{\arraystretch}{1.15}
    \begin{tabular}{|l|c|c|c|c|}
        \hline
         & $m_{\mathcal{G}}=25$ & $m_{\mathcal{G}}=50$ & $m_{\mathcal{G}}=100$ & $m_{\mathcal{G}}=200$ \\
        \hline
        Thm.4 
        & 0.0258  s
        & 0.2397  s
        &6.4194  s
        & 231.4300 s \\
        \hline
        Thm.3, $H=1000$
        & 0.0147   s
        &0.0318  s
        & 0.0627 s
        & 0.1109 s \\
        \hline
        Thm.3, $H=2000$
        & 0.0248 s
        &  0.0544 s
        & 0.0990 s
        & 0.1697 s \\
        \hline
        Thm.3, $H=4000$
        & 0.0421   s
        & 0.0961 s
        & 0.1620 s
        & 0.2514 s \\
        \hline
    \end{tabular}
\end{table}

As second step, we use Algorithm~\ref{alg:alternating} with a maximum iteration number $K = 5$ to train several recursive NFIR operators $\mathcal{F}_{\mathrm{NFIR}}(u,q,\theta_{\mathcal{G}},\theta_{\mathcal{N}})$ of size $m_{\mathcal{N}}= 10$, $R_{1}=R_{2}=1$, $m_{\mathcal{G}}=50$, for $q(t) =
    y(t-1) $ for all $t \in \mathbb{Z}$. As discussed in \eqref{eq:NN_for_lifting}, the nonlinear function in \eqref{eq:f} is represented by a MLP neural network with two hidden layers of dimensions $m_{W_{1}} = m_{W_{2}} = 4$, using the activation function $\sigma = \tanh$ in each layer. Therefore, the total number of identifiable parameters of the NFIR operators is always less than or equal to 
    \textcolor{black}{582.}
    We collect \(300\) batches of $\{(u_i,y_i)\}_{i=0}^{299}$ and construct $q_{i}$ from delayed $y_{i}$ to get 
$
\mathcal{S}=\{(u_i,q_i,y_i)\}_{i=0}^{299}
$. 

The minimum achieved by the algorithm shows an interesting dependence on the hyper-parameters 
$m_{\mathcal{G}}$ and $m_{\mathcal{N}}$. Changing the FIR order $m_{\mathcal{G}}$ from  $5$ to $50$ leads to a strong 
fitting improvement from $48.28 \%$ to $84.43\%$. However, moving from $50$ to $100$ produces only a modest further improvement, suggesting $m_{\mathcal{G}} = 50$ is a good choice.
This is interesting since $m_{\mathcal{G}} = 50$ corresponds to a finite impulse response length of 
\(\SI{1}{\second}\), which matches
the approximate \(\SI{1}{\second}\) time constant of 
\textcolor{black}{the plant.}
A similar behavior can be observed for the number of branches $m_{\mathcal{N}}$. Going from $1$ to $10$ leads to a fitting improvement from $41.75\%$ to $85.56\%$. Whereas going from $10$ to $20$ leads to a negligible further improvement.
Ultimately, increasing $m_{\mathcal{G}}$ and $m_{\mathcal{N}}$
provides a longer memory for the FIR filter and a wider set of branches, which guarantees additional expressivity but yields diminishing returns in performance gain after a certain threshold.
For the other hyper-parameters, for this particular example, we find that increasing the size of the memory of the nonlinear lifting operator, $R_{1}$ and $R_{2}$, or increasing the size of the layer of the neural network, $m_{W_{1}}$ and $m_{W_{2}}$, also lead to marginal improvements. For the computation, $m_{\mathcal{N}}$ is the key factor, with the computing time of Algorithm~\ref{alg:alternating} approximately doubling when $m_{\mathcal{N}}$ is increased from $20$ to $40$.  

The impact on the identification performance of the choice of the lifting operator, recursion, and passivity 
is verified using unseen test data,  
under both noise-free and noisy conditions (SNR = 10dB).
For a set of plant input-output (finite-length) data $(u,y) \in \ell_{2} \times \ell_{2}$, we consider the metric 
\begin{equation} \label{eq:fit_metric}
    \mathrm{Fit} [\%]
        =
        100\left(
        1 -
        \frac{
        \left\| y -\hat{y} \right\|_{2}
        }{
        \left\| y \right\|_{2}
        }
        \right)
\end{equation}
where $\hat{y}$ is the estimated model output, whose computation depends on the specific model structure under evaluation. This metric maps a normalized root mean squared error commonly used in system identification~\cite{Manchester2024TAC} to a percentage in the range $(-\infty,100]\%$, 
providing a convenient scale for comparing the relative magnitudes of different normalized errors.

\begin{figure}[htbp]
    \centering
    \vspace{-2mm}
    \includegraphics[width=1.03\columnwidth]{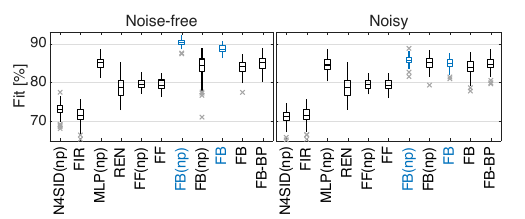}
    \vspace{-6mm}
    \caption{$\mathrm{Fit}$ performance on unseen test data for different model classes. Left: models trained on noise-free data. Right: models trained on noisy data. The suffix ``(np)'' denotes a non-passive model.}
    \label{fig:best_test_box}
\end{figure}

In Fig.~\ref{fig:best_test_box}, we compare our
identified NFIR operator with other operators. We denote our operator with \textbf{FB} and \textbf{FB(np)} where `FB' indicates the presence of feedback and `np' stands for no-passivity constraints.
For the other operators we consider
\textbf{N4SID(np)} - A LTI state-space model of order $20$, identified using MATLAB command $\mathrm{n4sid}$. This serves as the best possible linear model; 
\textbf{FIR} - A linear passive FIR filter with $m_{\mathcal{G}}=50$. This allows us to assess the impact of the nonlinear lifting operator;
\textbf{MLP(np)} -
A standard MLP neural network, with 631 parameters. 
The MLP uses an input `memory' of $50$ units,
that is, its input at time $t$ is $[u(t),u(t-1),\dots,u(t-49)]^{T} \in \mathbb{R}^{50}$. This way,
its input memory is comparable to that of FIR filters 
of our NFIR operator; 
\textbf{REN} - A passive recurrent equilibrium network of 702 parameters~\cite{Manchester2024TAC};
\textbf{FF} and \textbf{FF(np)} - A non-recursive NFIR, with the additional input $q$ removed. As above, `np' indicates no passivity constraints enforced. This comparison allows us to assess the impact of recursion and passivity on the identification performance;
\textbf{FB-BP} -
As discussed in Section~\ref{sec:simulation_error}, we trained an NFIR with a step of backpropagation applied in the last iteration of Algorithm~\ref{alg:alternating}.

Fig.~\ref{fig:best_test_box} shows the box plots of $\mathrm{Fit}$ for 100 batches of unseen test data. 
The NFIR appears to achieve a level of performance comparable to other models. 
Remarkably, this is true also for 
\textbf{MLP(np)} and \textbf{REN} networks, which
are universal approximators. 
Lifting and recursion definitively enhance expressivity. 
For this particular example, the impact of passivity on  performance is marginal. Furthermore, the sensitivity of our NFIR operator to noise appears similar to the one of other, well-established linear and nonlinear models. 

Two additional performances are shown in blue in Fig.~\ref{fig:best_test_box}, 
for both FB and FB(np) cases.
These capture the open-loop case, 
where the feedback signal 
\textcolor{black}{$q(t) =  \hat{y}(t-1) $}
for all $t \in \mathbb{Z}$
is replaced by the plant output
\textcolor{myblue}{$q(t) = y(t-1)$}
for all $t \in \mathbb{Z}$.
These additional evaluations highlight
the loss of performance due to the imprecise reconstruction of $\hat{y}$. Looking at FB-BP, it 
also shows how this issue can be alleviated with 
a final step of
 backpropagation through time in Algorithm \ref{alg:alternating} \footnote{The code of all examples is available at \url{https://github.com/Cambridge-Control-Lab/Passive_lifted_FIR_filters_for_nonlinear_system_identification}}.

\vspace*{-3mm}

\subsection{Industrial robot (experimental data)}
We  test the NFIR operator on real-world scenarios while assessing its capability when training is limited to 
small  data.  
We consider the industrial robot dataset introduced in~\cite{Wernholt2006}
and available as part of the MATLAB System Identification Toolbox~\cite{MatlabGreybox2,Matlab_NN_docu2}.
The dataset consists of four batches of multi-sine input–output data (collocated torque-velocity) sampled at 2\,$\mathrm{kHz}$.
In~\cite{MatlabGreybox2} these are denoted as \texttt{ue}, \texttt{uv1}, \texttt{uv2}, and \texttt{uv3}. Following~\cite{Matlab_NN_docu2}, we downsample the data to 200\,$\mathrm{Hz}$ so that each contains $1984$ time samples. We train only on the single $1984$-sample batch \texttt{ue} and test on the other three batches, whereas the previous example used significantly more data ($300$ training batches with $75{,}000$ samples in total). 

We compare (i) a passive linear FIR operator with $m_{\mathcal{G}}=500$, (ii) the state-of-the-art passive nonlinear grey-box model~\cite{MatlabGreybox2}, based on a three-mass nonlinear flexible robot arm with physical parameters, and (iii) a passive NFIR operator. The NFIR operator is constructed this time as the parallel combination of a linear FIR filter and a small NFIR operator, capturing the `residual' nonlinear phenomena. The latter has the same structure of Section \ref{example:mass-spring},
with hyper-parameters $R_{1}=R_{2}=150$, $m_{\mathcal{G}}=500$, $m_{W_{1}} = m_{W_{2}} = 8$ and $m_{\mathcal{N}}=3$. 
The total number of identifiable parameters is $4507$. 

We set the maximum iteration in Algorithm~\ref{alg:alternating} to $K=4$.
We find that increasing the regularization $\gamma_{\mathcal{G}}, \gamma_{\mathcal{N}}$ in~\eqref{eq:rand1} and using a small  $m_{\mathcal{N}}$  reduce overfitting.  
Results are summarized in Table~\ref{tab:industrial_robot}. 
The NFIR does better than the grey-box model on 
training and testing batches \texttt{ue} and  \texttt{uv1}. This is probably due to the fact that \texttt{uv1} has a waveform similar to \texttt{ue}, as shown in~\cite{MatlabGreybox2}. Although \texttt{uv2} and \texttt{uv3} have significantly different waveforms from \texttt{ue} (see~\cite{MatlabGreybox2}), the NFIR performance degrades but remains  comparable to the one of the grey-box model. 
The degradation on \texttt{uv2} suggests that \texttt{ue} may provide an insufficiently exciting input, since even the grey-box model performance degrades. 
  
\begin{table}[htbp]
    \centering
    \caption{Industrial robot model-fit performance.}
    \label{tab:industrial_robot}
    \setlength{\tabcolsep}{5pt}
    \renewcommand{\arraystretch}{1.12}
    \begin{tabular}{lcccc}
        \toprule
         & \texttt{ue} & \texttt{uv1}  & \texttt{uv2}  & \texttt{uv3}  \\
        \midrule
        FIR
        & \textbf{85.19\%} & 85.54\% & 47.60\% & 86.17\% \\
        
        Grey-box
        & \textbf{85.60\%} & 85.26\% & 57.68\% & 95.55\% \\
        
        NFIR
        & \textbf{92.29\%} & 89.88\% & 57.34\% & 83.14\% \\
        \bottomrule
    \end{tabular}
\end{table}

\section{Comparison with the literature} \label{sec:related_work}    
      When the lifting operator $\mathcal{N}$ is given by a finite number of kernel sections and the linear filter $\mathcal{G}$ is represented  as a finite-dimensional matrix,
      NFIR operators can be interpreted as a restricted causal version of the nonlinear operators proposed in \cite{Henk2024}.
    Even with these restrictions, our NFIR operators extend the operators of \cite{Henk2024} by including recursion and by addressing the issue of causality, which is essential for real-time implementation.
     Passivity in \cite{Henk2024} is established via LMI constraints, which here we replace with more efficient frequency-domain constraints.
   
    Incrementally passive operators within reproducing kernel Hilbert spaces have been studied in \cite{Henk2023} and \cite{Tony2024}.
    In \cite{Henk2023}, incrementally passive 
    operators are obtained by scattering transform
    from identified operators that have 
     incremental gain less than one. The latter are derived
     via regularized unconstrained optimization. \cite{Tony2024} further refines these results, 
     through the identification of memory functionals. These allow for the derivation of causal operators as in this paper. A key distinction with our work is that we
     enforce  passivity directly, via constrained optimization. Passivity is also a weaker property than incremental passivity. This difference is significant. As discussed in \cite[Sec. IX.B]{Henk2023} and \cite[Sec. I]{Henk2024}, the use of the scattering transform to derive passive but not incrementally passive operators is an open problem.

    The proofs of Theorems \ref{thm:lift_operator} 
    and \ref{thm:NFIR_passive} leverage the lifting of linear passive FIR filters into a larger feature space, based on a pre- and post-composition with a 
    lifting operator and its adjoint, both carefully designed to ensure the causality of the resultant nonlinear operator. This idea has been considered in \cite[Prop. 2.11 and Chp. 4.3]{Rodolphe1996} for
    state-space systems. In this paper, we have developed this lifting within the operator-theoretic setting while providing the optimization-based formulation to address identification problems.

\section{Conclusion}
We have introduced the class of passive NFIR operators, which are linear passive FIR filters in a lifted space induced by nonlinear lifting functions. We proposed an alternating algorithm that jointly learns the lifted passive FIR filters and the nonlinear lifting function. For learning FIR filters, we use convex optimization and enforce passivity based on frequency sampling. Enforcing passivity is a key contribution of this approach, which holds true irrespective of the quality and quantity of the training data. 

This work opens several promising research directions. The parallel bank of FIR filters at the core of NFIR operators could be generalized to richer structures. 
For instance, we could adopt linear infinite impulse response (IIR) filters in both continuous and discrete time. Such developments would broaden the applicability of the framework and enable analog implementations. Other promising avenues are the extension to data-driven nonlinear  control, starting from revising \cite{Wang2024} within a nonlinear setting, and the real-time adaptation of the lifting operator, which could open new directions in adaptive control theory (within an operator-theoretic setting). These research questions are beyond the scope of this paper and will be addressed in future works. \vspace{-2mm}
\bibliographystyle{IEEEtran} 
\bibliography{bib_edit_abbrev,IEEEabrv}
\end{document}